\documentclass[%
 reprint,
 amsmath,amssymb,
 aps,
 pra,
]{revtex4-1}

\usepackage{graphicx}
\usepackage{dcolumn}
\usepackage{bm}

\usepackage{amsfonts}
\usepackage{dcolumn}
\usepackage{bm}
\usepackage{todonotes} \usepackage{setspace}

\newcommand{\Ai}[1] {{\rm Ai}\hspace{-0.1cm}\left(#1\right)}

\newcommand{\Bi}[1] {{\rm Bi}\hspace{-0.1cm}\left(#1\right)}

\newcommand{\AiH}[2] {{\rm Ai}H^{(#1)}\hspace{-0.1cm}\left(#2\right)}

\begin{document}

\title{Full characterisation of Airy beams under physical principles}

\author{ J Rogel-Salazar}
\email{j.rogel@physics.org}
\affiliation{Applied Mathematics and
  Quantitative Analysis Group, Science and Technology Research
  Institute, School of Physics Astronomy and Mathematics, University
  of Hertfordshire, Hatfield, AL10 9AB, U.K.} 
\altaffiliation{Blackett Laboratory, Department of Physics, Imperial
  College London, Prince Consort Road, London, SW7 2BZ, UK } 

\author{H A Jim\'enez-Romero}
\affiliation{Escuela
  Superior de F\'{\i}sica y Matem\'aticas, Instituto Polit\'ecnico
  Nacional, Edificio 9, U. P. Adolfo L\'opez Mateos, M\'exico,
  D. F. 07300, M\'exico } 

\author{S Ch\'avez-Cerda} 
\email{sabino@inaoep.mx}
\affiliation{Group of Theoretical Optics,
  Instituto Nacional de Astrof\'{\i}sica, \'Optica y Electr\'onica,
  Apartado Postal 51/216, 72000, Puebla, Puebla, M\'exico}

\begin{abstract}
  The propagation characteristics of Airy beams is investigated and
  fully described under the traveling waves approach analogous to that
  used for non-diffracting Bessel beams. This is possible when
  noticing that Airy functions are in fact Bessel functions of
  fractional order $\frac{1}{3}$. We show how physical principles
  impose restrictions such that the non-diffracting Airy beams cannot
  be of infinite extent as has been argued, and introduce for the
  first time quantitative expressions for the maximum transverse and
  longitudinal extent of Airy beams. We show that under the
  appropriate physical conditions it is possible to obtain
  higher-order Airy beams.
  
\end{abstract}



\maketitle

Bessel beams are known to belong to a class of optical beams described
by the $(2+1)$-dimensional Helmholtz equation, and they have the
property of being non-diffracting \cite{Durnin1987, Chavez99} and
self-healing \cite{Bouchal1998207, Anguiano-Morales2007} on
propagation. Originally, Bessel beam properties have been studied
using the diffraction Rayleigh-Sommerfeld or Fresnel-Kirchhoff
integrals. Implicitly, this approach considers the Bessel beam to be
of infinite transverse extent \cite{Durnin1987}. Alternatively, a more
straightforward and physically sound description of these beams can be
done in terms of travelling Hankel waves \cite{Chavez99}. This
description based on the differential Helmholtz wave equation provides
a better understanding of the physical origin of these beams and their
intriguing properties. Even further, this same approach allowed to
demonstrate the existence of other families of non-diffracting beams
described by fundamental solutions of the separable (2+1) Helmholtz
wave equation. These families of beams show propagation
characteristics analogous to the Bessel beams \cite{mathieu,parab}.

In recent years, another kind of non-diffracting beam has been
reported but its propagation is governed by the paraxial wave equation
in $(1+1)$ dimensions, the so-called Airy beams
\cite{Siviloglou07}. These beams do not belong to the same class as
those of the Helmholtz equation and thus it might be expected that
their physical properties are not actually described in the same
terms. Nonetheless, an interesting fact that is hardly discussed in
all the published literature on Airy beams is that the Airy functions
are Bessel functions of fractional order equal to $1/3$. This allows
the application of the aforementioned travelling Hankel wave
description. In this work we show that Airy beams have similar
properties to those of Bessel beams due to the fact that the former
are the result of the superposition of counter-propagating Hankel
travelling waves of fractional order. We also demonstrate that the
proposed wave approach imposes the condition under physical principles
for these beams to be of finite extent contrary to the ``ideal''
infinite Airy beam.

The aim of this paper is to provide a better understanding of the
fundamental nature of the travelling wave methodology applied to Airy
beams. In particular we show that all the known propagation
characteristics of Airy beams are straightforwardly and intuitively
understood using the Hankel travelling wave approach and that the
focusing features of $(1+1)$-dimensional Airy beams can only be
described in clear and simple terms with this methodology, which is
similar to that for $(2+1)$-dimensional Bessel beams. In Section
\ref{sec:Hankel} we provide a brief account of the travelling wave
description for Bessel beams as a way to motivate its application to
Airy beams. Section \ref{sec:TravellingAiry} explains the relationship
between Airy and Bessel functions and in Section \ref{sec:finite} we
show the characterise of the finite Airy beam in a direct
manner. Finally in Section \ref{sec:HigherOrder} we show that there
can exist higher order Airy beams by discussing the physical
conditions that enable this possibility.

\section{Travelling Hankel Waves and Bessel Beams}
\label{sec:Hankel}
Light propagation in linear media is described with the use of the
scalar wave equation for the electric field $E({\bf r},t)$ given by
\begin{equation}
  \label{eq:ScalarWaveEq}
  \nabla^2 E({\bf r},t)=\frac{1}{v^2}\frac{\partial E({\bf
      r},t)}{\partial t^2}.
\end{equation}
In cylindrical coordinates ${\bf r}=(r,z)$, and $v$ corresponds to the
speed of light in the medium in question. It is possible to solve
Equation (\ref{eq:ScalarWaveEq}) by separation of variables
\cite{boas2005mathematical} ending up with ordinary differential
equations for the variables $r$, $z$ and $t$. The separation constants
can be chosen such that $\omega^2/v^2=k_r^2+k_z^2\equiv |{\bf k}|^2$
so that we can think of ${\bf k}$ as the wavevector. In this manner,
the equation for the radial coordinate from Equation
(\ref{eq:ScalarWaveEq}) is given by:
\begin{equation}
  \label{eq:BesselEq}
  \frac{d^2 H}{d r^2}+ \frac{1}{r}\frac{d H}{d r}+
  \left(k_r^2 - \frac{m^2}{r^2}\right)H=0,
\end{equation}
which is the Bessel differential equation of order $m$
\cite{watson1995treatise} and its solutions are given by the $m-$th
order Bessel function, $J_m(k_r r)$ and the $m-$th order Neumann
function, $N_m(k_r r)$. In many cases the Neumann functions mentioned
above are discarded due to the singularity they present at the
origin. However, it has been proved by one of the authors
\cite{Chavez99} that these functions do indeed carry physical
meaning. Also, the solutions $J_m$ and $N_m$ cannot be used separately
to describe the \emph{propagation} of light, because they do not
satisfy the Sommerfeld radiation condition independently. For
cylindrical waves this condition reads,
\begin{equation}
  \lim_{r\rightarrow \infty} r^{1/2} \left(\frac{dH}{dr}-ik_rH \right)=0,
  \label{eq:Sommerfeld}
\end{equation}
and it tells us that a wave equation cannot have waves coming from an
infinite distance. Thus, the solution needed to describe
\emph{propagating} waves must be given by the complex superposition of
the $J_m$ and $N_m$ functions. This leads to the so-called Hankel
waves \cite{Chavez99}
\begin{eqnarray}
  \label{eq:H1}
  H_m^{(1)}(k_r r)= J_m(k_r r) + i N_m(k_r r),\\
  \label{eq:H2}
  H_m^{(2)}(k_r r)= J_m(k_r r) -  i N_m(k_r r).
\end{eqnarray}
Once the azimuthal and longitudinal wave components are incorporated, $\exp(i m \varphi+ik_z z)$, 
the wavefronts of the solutions for the Helmholtz wave equation in
cylindrical coordinates are conic helicoids. $H_m^{(1)}$ and $
H_m^{(2)}$ are related to outgoing (convex) and incoming (concave)
conic solutions respectively. The Hankel waves, in combination with
the temporal part, describe cylindrical wavefronts that collapse and
are generated at the longitudinal $z$-axis. This is the physical
origin of the singularity of the Hankel functions, the longitudinal
axis is simultaneously sink and source for the incoming and outgoing
cylindrical wave components, respectively. In that sense, since
incoming waves become outgoing, there is a region where both of them
interfere leaving only the Bessel function:
\begin{eqnarray}
  E_{in}(r,\varphi,z,t)+&&E_{out}(r,\varphi,z,t)=\nonumber \\
&&2J_m(k_r r)\exp\left(im\varphi + ik_z z - i \omega
    t\right).
  \label{eq:2}
\end{eqnarray}
Recalling the Sommerfeld radiation condition, Equation
(\ref{eq:Sommerfeld}), the incoming wave must be generated at a finite
distance implying that Bessel beams must have finite transverse extent
and thus a finite propagation distance. The interference of both
Hankel waves only occurs within a conic region, and it is only within
this region that the Bessel beam can be formed. This region is
therefore called {\it the region of existence} of the Bessel beam
which can be exploited in applications such as the design of laser
resonators whose output is related to a Bessel beam
\cite{Rogel-Salazar2001}.

As we have mentioned before, Bessel beams are non-diffractive,
i.e. they propagate without spreading or changing their shape within
their region of existence. Similarly, they show the property of
self-healing that occurs when the beam is partially blocked, i.e. it
reconstructs itself after some distance
\cite{Anguiano-Morales2007}. These properties can straightforwardly be
understood in terms of travelling Hankel waves that provides a clear
frame for the physics behind of them and others like the evolution of
focused Bessel beams \cite{Chavez00}.

\section{Travelling Waves Approach to Airy Beams}
\label{sec:TravellingAiry}
Berry and Balazs introduced the idea of the ``non-spreading Airy
wave-packet'' by solving a force-free Schr\"odinger equation
\cite{Berry79}. The solution propagates without change along a
parabolic trajectory and thus show acceleration. The relationship
between the force-free Schr\"odinger equation and the paraxial wave
equation indicates that non-diffractive Airy beams are therefore
possible and indeed they have been observed
\cite{Siviloglou07}. Furthermore, it has been reported that, similar
to Bessel beams, Airy beams also have the property of self-healing
\cite{Broky08}.

Airy beams can be obtained when considering the propagation of a
plane-polarised beam in a linear medium described by the normalised
paraxial wave equation
\begin{equation}
  \label{eq:ParaxialWaveEq}
  -i\frac{\partial U}{\partial \xi}+\frac{1}{2}\frac{\partial^2
    U}{\partial s^2}=0,
\end{equation}
where $U(s, \xi)$ is the electric field envelope that depends on the
normalised coordinates $s=x/x_0$ and $\xi=z/kx_0^2$, with $x_0$ being
a given transverse scale and $k$ is the wavenumber. Since we know that
the Airy beam remains invariant while propagating along a curved
trajectory we can define an accelerating variable
$\Gamma=s-\frac{a}{4} \xi^2+ v \xi$ with $a$ and $v$ being real
constants. We can now write the electric field envelope as follows
$U(s,\xi)=w(\Gamma)\exp(i\theta(\Gamma,\xi))$, which leads to the
following ordinary differential equation for $w(\Gamma)$
\begin{equation}
  \label{eq:w}
  w''-\alpha \Gamma w = \frac{A^2}{w^3},
\end{equation}
and $\theta(\Gamma,\xi)$ is given by
\begin{equation}
  \label{eq:Theta}
  \theta(\Gamma,
  \xi)=A\int_0^\Gamma\frac{d\Gamma'}{w^2}+\left(\frac{a}{2}\xi -
    v \right) + \frac{a^2}{24}\xi^3-\frac{a
    v}{4}\xi^2 + \frac{v}{2}\xi,
\end{equation}
where $A=A(\xi)$ is assumed to be a constant. We can think of the
parameters $a$ and $v$ above as the acceleration and velocity of the
beam, respectively. Without loss of generality, let us consider the
case where $a=1$, $v=0$ and $A=0$; Equation (\ref{eq:w}) becomes:
\begin{equation}
  \label{eq:AiryEq1}
  w''-\left(s-\frac{1}{4}\xi^2 \right)w=0,
\end{equation}
which is the well-known Airy differential equation \cite{Abramowitz}
and therefore the solution can be expressed as:
\begin{equation}
  \label{eq:SolAiry}
  U(s,\xi)=\Ai{s-\frac{\xi^2}{4}}\exp\left[
    i\left(\frac{s\xi}{2}-\frac{\xi^3}{12} \right)\right]. 
\end{equation}
where $\Ai{\cdot}$ is the Airy function. The expression above implies
that the intensity of the beam has the profile of the modulus squared
of the Airy function, i.e.
\begin{equation}
  \label{eq:Intensity_Airy}
  I(s,\xi)=\left|\Ai{s-\frac{\xi^2}{4}}\right|^2.
\end{equation}
The argument of the Airy function in Equation (\ref{eq:SolAiry}) shows
that the Airy beam follows a parabolic trajectory that can be
interpreted as having a transverse acceleration \cite{Berry79}. It is
also clear that it does not change neither its profile nor its
amplitude on propagation when observed along this parabolic
trajectory. In this sense the Airy beam can be regarded as being a
non-diffracting beam, as it is the case for the Bessel beam discussed
in Section \ref{sec:Hankel}. It may seem that although both Airy and
Bessel beams are non-diffracting, the physics are not described in a
similar way given that they do not belong to the same class of
beams. However, in what follows we will show that Airy beams can be
better understood when described by the travelling waves approach used
in the treatment of Bessel beams outlined in Section
\ref{sec:Hankel}.

In order to elucidate the properties and behaviour of Airy beams, let
us recall Airy's differential equation, namely
\begin{equation}
  \label{eq:Airy_Eq}
  \frac{d^2 w}{ds^2}\mp sw=0,
\end{equation}
whose solutions are given by the Airy functions $\Ai{s}$ and $\Bi{s}$
\cite{Abramowitz}, where we are using normalised coordinates. The
$\Bi{s}$ function is defined as the solution with the same amplitude
of oscillation as $\Ai{s}$ as $s$ goes to minus infinity and differs
in phase by $\pi/2$. We will therefore concentrate on the behaviour of
$\Ai{s}$ in the rest of this discussion. The Airy differential
equation is defined in the entire real space and thus, the second term
in Equation (\ref{eq:Airy_Eq}) can have either a positive or a
negative sign depending on whether the value of $s$ is positive or
negative. Using either of the two signs yields the same Airy solution
with the only difference that one will be the mirror reflection of the
other with respect to the vertical axis, $s=0$.

\begin{figure}[htbp]
  \centering
  \includegraphics[width=9cm]{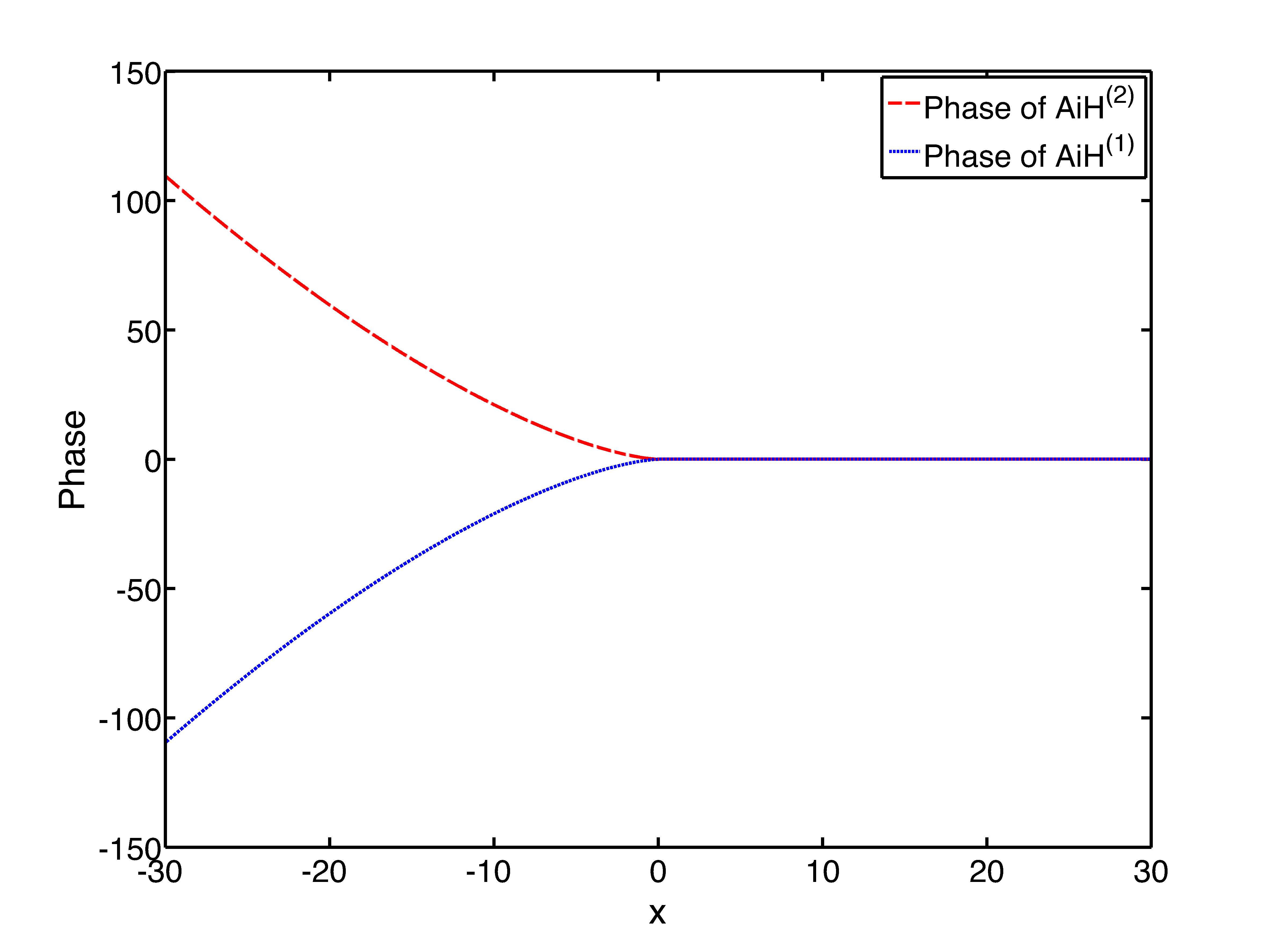}
  \caption{(Color online) Schematic wavefronts of the $\AiH{1}{s}$ and
    $\AiH{2}{s}$ functions (arbitrary units).}
  \label{fig:Phases}
\end{figure}

A simple calculation shows that by making the change of variable
$w=\sqrt{s} Z_{1/3} \left(\frac{2}{3} s^{3/2}\right)$, Equation
(\ref{eq:Airy_Eq}) can be transformed into the Bessel differential
equation of order $1/3$, with $Z_{1/3}$ being the cylindrical Bessel
function of order $1/3$ \cite{Gradshteyn, Jahnke43}. In a similar way,
the Airy functions can be expressed in terms of modified Bessel
functions of order $1/3$, $K_{1/3}$, as follows
\cite{Abramowitz}: 
\begin{eqnarray}
  \label{eq:w_Airy}
  w(s)&=&\frac{1}{\pi}\sqrt{\frac{s}{3}} K_{1/3}\left(\frac{2}{3}s^{3/2}\right),\\
  \label{eq:w_i_Airy}
  w(s)&=&\frac{1}{2}\sqrt{\frac{s}{3}} e^{2\pi i/3}H^{(1)}_{1/3}\left(\frac{2}{3}is^{3/2}\right).
\end{eqnarray}
We can distinguish two important cases; let us consider the
argument of these Bessel functions, $K_{1/3}$ to be
$\chi=\frac{2}{3}|s|^{3/2}$. On the one hand, when $s\ge 0$, the
profile has a monotonic decreasing behaviour and is proportional to
the modified Bessel function $K_{1/3}$:
\begin{equation}
  \label{eq:Airy_K}
  \Ai{s}=\frac{1}{\pi}\sqrt{\frac{s}{3}}K_{1/3}(\chi).
\end{equation}
On the other hand, when $s<0$, the profile can be considered as the
superposition of two waves which are essentially Hankel functions of
order $1/3$, i.e.
\begin{equation}
  \label{eq:Airy_H}
  \Ai{-s}=\frac{1}{2}\sqrt{\frac{s}{3}}\left[e^{\frac{i\pi}{6}}
    H_{1/3}^{(1)}(\chi) + e^{\frac{-i\pi}{6}} H_{1/3}^{(2)}(\chi)\right].
\end{equation}
We can now define the following functions:
\begin{eqnarray}
  \label{eq:AiH1}
  \AiH{1}{s}&=&\frac{1}{2}\sqrt{\frac{s}{3}} e^{\frac{i\pi}{6}} H_{1/3}^{(1)}(\chi),\\
  \label{eq:AiH2}
  \AiH{2}{s}&=&\frac{1}{2}\sqrt{\frac{s}{3}} e^{\frac{-i\pi}{6}} H_{1/3}^{(2)}(\chi).
\end{eqnarray}
With these definitions, we can now write the Airy function in a more
compact form as follows:
\begin{equation}
  \label{eq:Airy_compact}
  \Ai{-s}=\AiH{1}{s} + \AiH{2}{s}.
\end{equation}
We identify these functions as the travelling Hankel components of the
Airy beam, in analogy to the way it was done for Bessel beams in
Section \ref{sec:Hankel}.  We note that it is possible to obtain an
analytical form of the phase of these Hankel components by considering
the asymptotic expansion of Hankel functions, which approximate very
well to the original function from the first maximum
\cite{Abramowitz}. Using this expansion we have that
\begin{eqnarray}
  \label{eq:AiH1_expansion}
  \AiH{1}{s}&\simeq&\sqrt{\frac{s}{6\pi\chi}} e^{i\phi},\\
  \label{eq:AiH2_expansion}
  \AiH{2}{s}&\simeq&\sqrt{\frac{s}{6\pi\chi}} e^{-i\phi},
\end{eqnarray}
with $\phi=\chi-\frac{\pi}{4}$, and their wavefront being the opposite
of each other. In Figure \ref{fig:Phases}, we can see the wavefronts
of each of the components $\AiH{1}{s}$ and $\AiH{2}{s}$.
Each phase determines a geometric wavefront, and thus according to
geometrical optics, the rays that make up each beam will propagate
perpendicularly to this front \cite{Born}.  In fact, when the Airy
beam is cut, it will propagate along trajectories determined by the
rays of the $\AiH{2}{s}$ beam. We note that these waves must satisfy
the Sommerfeld radiation condition that imposes the restriction of
Airy beams being of finite extent and, similar to Bessel beams, they
can only exist within a finite region of space. 

\begin{figure}[htbp]
  \centering
  \includegraphics[width=9cm]{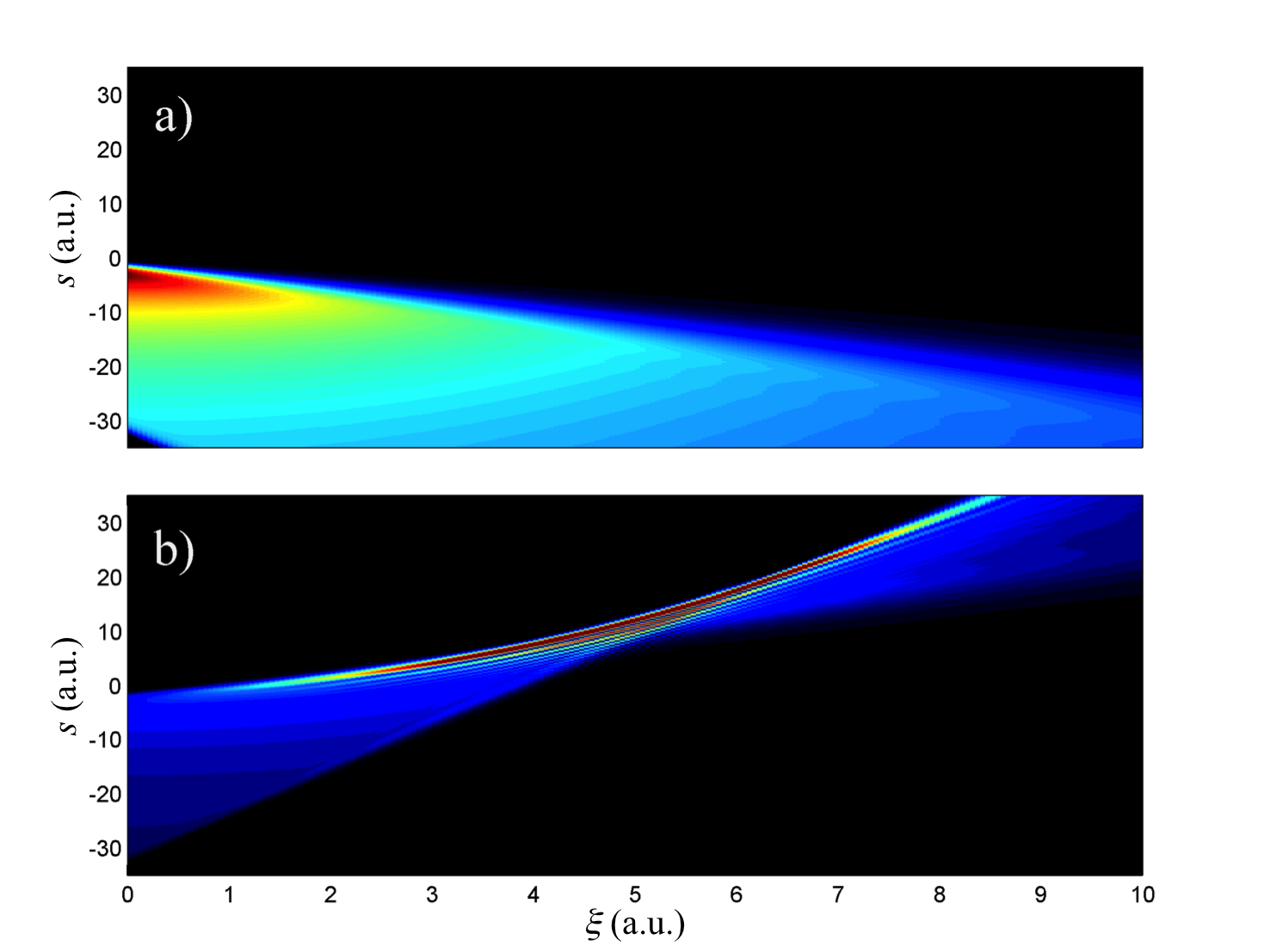}
  \caption{(Color online) Behaviour of the independent propagation of
    the ${\rm Ai}H^{(1)}$, a), and of the ${\rm Ai}H^{(2)}$, b).
    component of the Airy beam in arbitrary dimensionless units.}
  \label{fig:Focus_no_AiH1}
\end{figure}

In order to show that it is in fact the ${\rm  Ai}H^{(2)}$ component the one that bears the property of the parabolic caustics we carried out the propagation of each component independently. In Figure \ref{fig:Focus_no_AiH1}a we see the propagation of the ${\rm Ai}H^{(1)}$ component that simply travels away from the propagation axis diminishing its amplitude. In Figure
\ref{fig:Focus_no_AiH1}b the parabolic caustic associated to the
normal rays to its concave wavefront can indeed be appreciated. We note that the caustic determined by the rays defines the parabolic
motion of the main maximum of the beam under consideration
\cite{Berry79}. This behaviour is similar to that produced by a third order aberration (coma). The numerical simulations shown in this paper use a
computational window such that the dimensionless parameters are $s\in[-30,30]$ and $\xi\in[0,5]$, and we have used a super-Gaussian window
$t(x)=\exp{-(s/t_0)^{50}}$, $t_0$ the width of the window, to reduce
diffraction effects introduced if a hard aperture was used instead and
emphasize the propagation features of the AiH beams.

\begin{figure}[htbp]
  \centering
  \includegraphics[width=9cm]{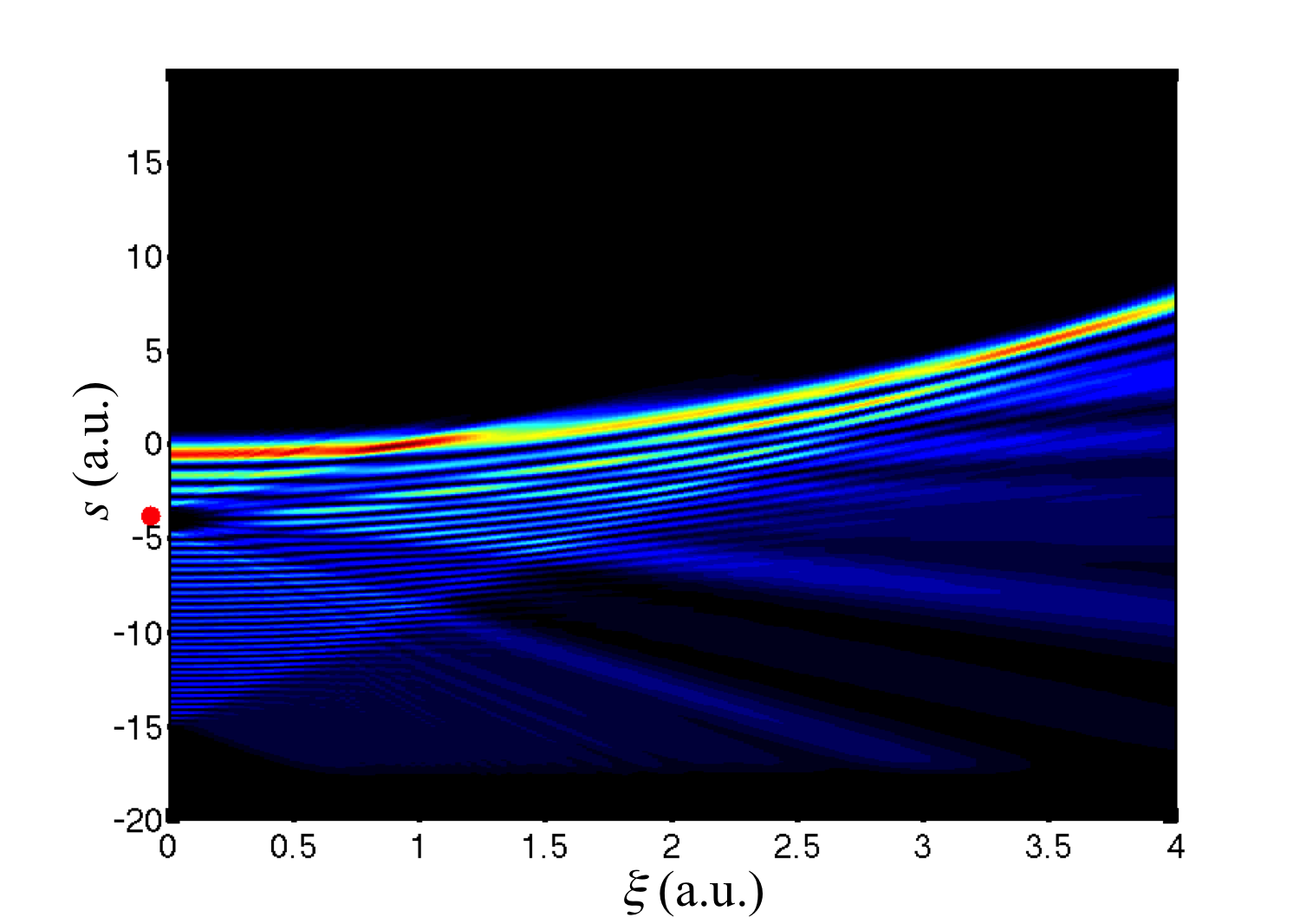}
  \caption{(Color online) Propagation of an obstructed Airy beam (in
    arbitrary dimensionless units) showing self-healing. Notice the
    presence of two shadows. The red dot at the edge of the figure
    indicates the position where the obstruction has been located.}
  \label{fig:Self_healing}
\end{figure}

We can now easily understand the self-healing property of the Airy
beam: since the Airy beam is formed by the superposition of the
fractional Hankel components $\AiH{1}{s}$ and $\AiH{2}{s}$, the
self-healing region arises from the recombination of these two
components after the shadows of the obstructed waves as seen in Figure
\ref{fig:Self_healing}.

\begin{figure}[htbp]
  \centering
  \includegraphics[width=8.5cm]{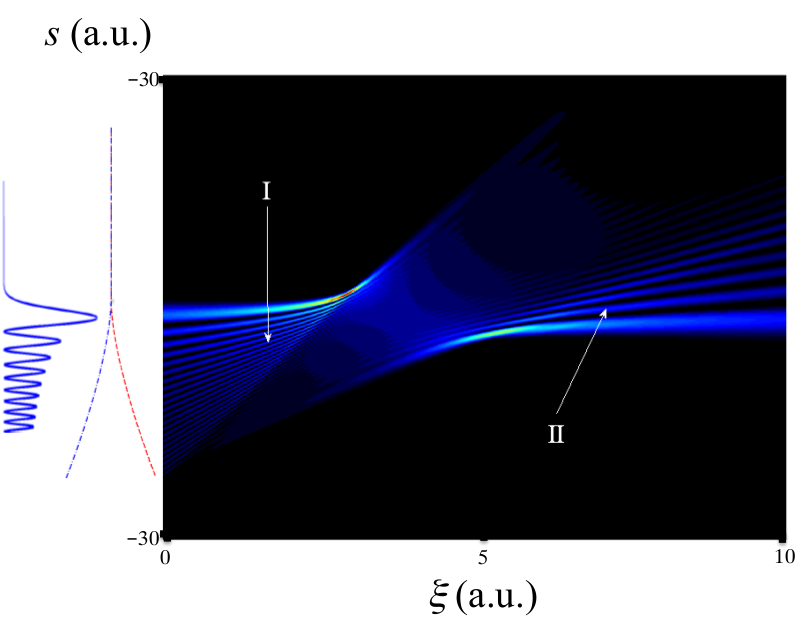}
  \caption{(Color online) Behaviour of a focussed Airy beam (in
    arbitrary dimensionless units), observe the two regions due to the
    focusing of the two Hankel travelling waves. In the edge of the
    figure we have indicated the profile of the Airy beam as well as
    the two wavefronts of the beam.}
  \label{fig:Double_Focusing}
\end{figure}

In regards to focusing, we note that the Airy beam shows the
non-common behaviour of presenting two different focusing regions, one
for $\AiH{2}{s}$ in which focusing is observed and a second for
$\AiH{1}{s}$, farther away, where a defocusing beam appears. This
behaviour can be seen in Figure \ref{fig:Double_Focusing}, where we
have marked the two regions as I and II. This behaviour is easily
explained by noting that the composing Hankel waves, besides having
opposite travelling directions, have opposite wave front curvatures:
one is positive focusing and the other is negative defocusing. When
passing through the positive lens these add or subtract accordingly.

\begin{figure}[htbp]
  \centering
  \includegraphics[width=8.5cm]{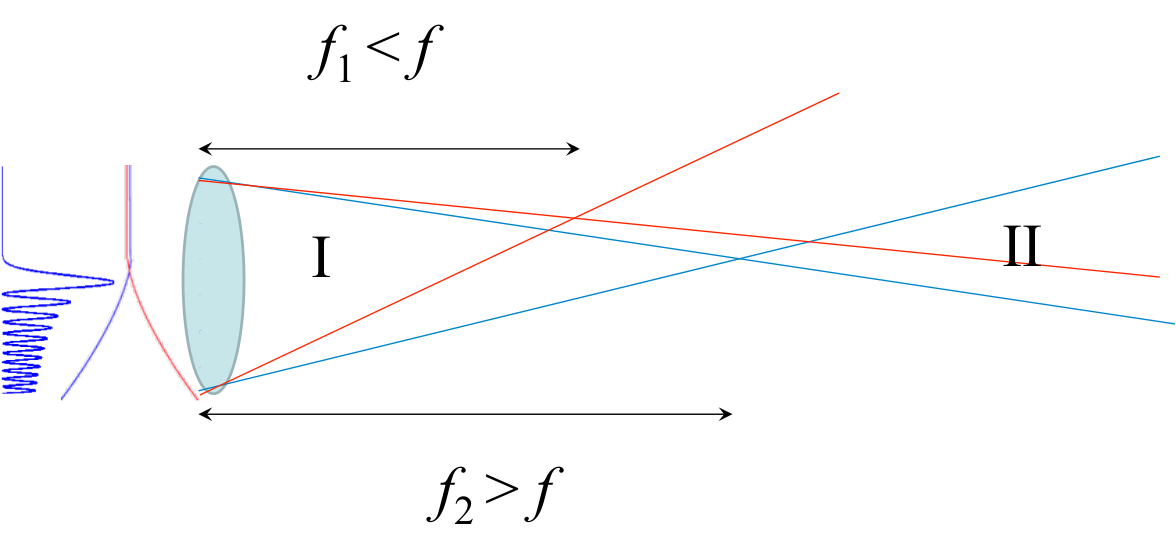}
  \caption{Schematic ray tracing of the double focusing of an Airy beam related
    to that shown in Figure \ref{fig:Double_Focusing}.}
  \label{fig:Double_Focusing_Ray}
\end{figure}

A simple geometrical analysis under this consideration easily explains
the two observed regions. In Figure \ref{fig:Double_Focusing_Ray} we
show a schematic of the optical arrangements including a lens; the
wavefronts have been marked in two different colours. Notice how each
region is generated by the focusing of the rays coming from each of
the two wavefronts, i.e. from each of the Hankel waves that compose
the Airy beam.
We can think of each region to be formed due to the simultaneous
incidence of each Hankel component of the Airy beam. We would like to
emphasise that this is the case regardless of any interference pattern
caused by the functions ${\rm Ai}H^{(1)}$ and ${\rm
  Ai}H^{(2)}$. However, we can indeed go further and assume that it is
possible to make use of a packet that is either only the real or only
the imaginary part of one of these two components, say ${\rm
  Ai}H^{(1)}$ for example.

\begin{figure}[htbp]
  \centering
  \includegraphics[width=9cm]{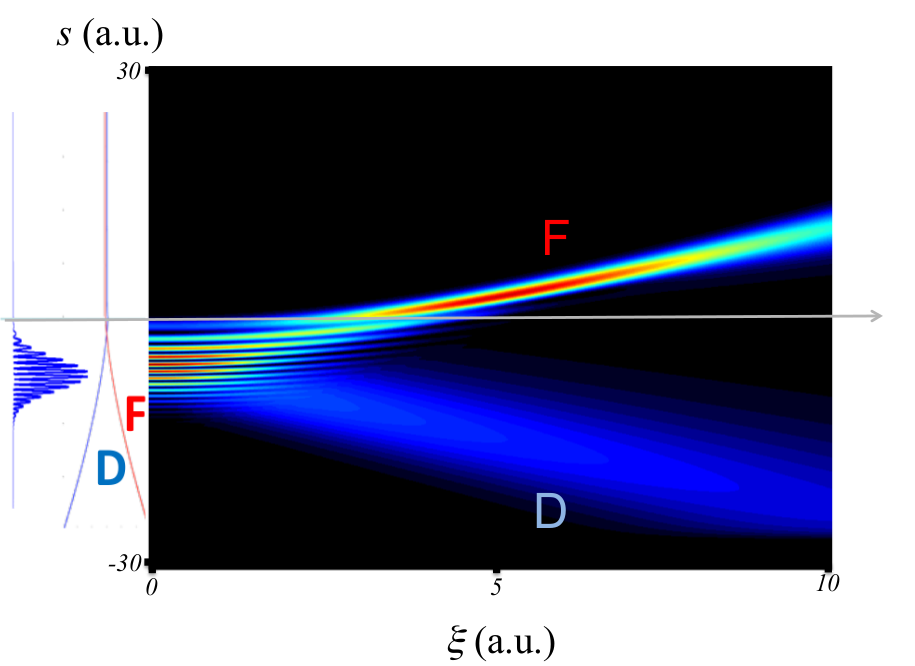}
  \caption{(Color online) Propagation of off-axis Airy-Gauss beam
    using arbitrary dimensionless units}
  \label{fig:Airy_Gauss}
\end{figure}

Finally, we can now take a more general view of the behaviour of an
Airy-Gauss beam when it is off-Axis. In Figure \ref{fig:Airy_Gauss} we
show the propagation of the Airy-Gauss beam, where it is clear that
each of the Hankel components provides the two contrasting behaviours
of focusing and defocusing mentioned before.

As we have seen, the application of the travelling Hankel wave
approach has provided us with a straightforward description of the
propagation characteristics of the Airy beam under different
circumstances. We will now show how to determine the propagation
distance of finite energy Airy beams.

\section{Finite Airy beams}
\label{sec:finite}
It is usually argued that non-diffracting beams, such as the Airy
beam, require an infinite amount of energy for their generation. This
might be to give an explanation for the non-diffraction feature or
because the range of the mathematical function that describes the
profile is infinite. However, in this Section we show that from a
physical perspective this cannot be the case as the beam must be
constrained by physical principles. One of them was mentioned above
for the Bessel beams that also apply for the Airy beams and this is
the Sommerfeld radiation condition. Within the travelling wave
approach, to have an infinite Airy beam would require having sources
at infinity.

For Airy beams, whose profile exhibits reduction of the separation distance between intensity peaks, there exists another more basic physical constraint that we discuss now. The distance between two consecutive peaks in the transverse Airy intensity profile should not become smaller than the wavelength of the light used. If that were the case we would end up
with an unphysical situation. This situation is analogous to the
treatment, for instance, of wave excitations with a fractal boundary
\cite{Sapoval1989,Sapoval1991}; although mathematically the fractal
structure continues to infinitely small scales, in reality there are
physical constraints that avoid this situation.

\begin{figure}[htbp]
  \centering
  \includegraphics[width=9cm]{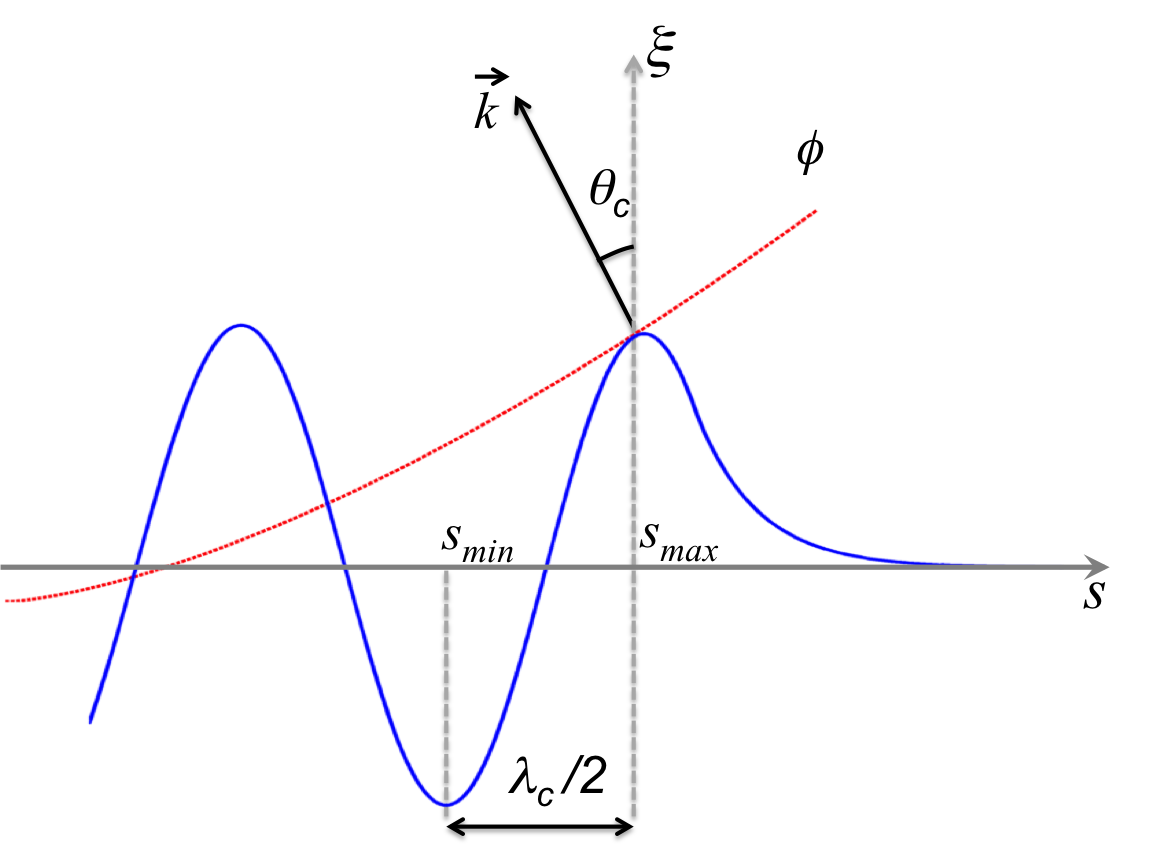}
  \caption{(Color online) Schematic of the behavior of an Airy beam:
    When the wavevector $\protect\overrightarrow k$ of the Airy beam,
    which is normal to the wavefront $\phi$, reaches the maximum
    paraxial value $\theta_c=\frac{\pi}{6}$ the oscillations in the
    profile must dampen in order to remain physical.  At that point,
    the distance between two extreme points in the profile must be of
    the order of half the critical wavelength $\lambda_c$.}
  \label{fig:AiryDamping}
\end{figure}

Next, we proceed to find a critical value of the beam extent after
which its profile must dampen, giving rise to a finite energy Airy
beam. In order to tackle the issue, let us take the asymptotic
expansion for large arguments of the Airy function given by the
corresponding Bessel functions of order $1/3$:
\begin{equation}
  \label{eq:Hankel_expansion}
  \Ai{\alpha s}\propto \cos\left(\frac{2}{3}(\alpha s)^{\frac{3}{2}}-\frac{\pi}{4} \right),
\end{equation}
where we have taken the normalised coordinates and $\alpha$ is a
parameter that allows us to change the frequency in the Airy
function. We need to find the distance between two consecutive extreme
points of the equation above, see Figure \ref{fig:AiryDamping}. These
points occur when the following conditions are met: For minima we have
\begin{equation}
  \label{eq:Smin}
  s_{min}=\frac{1}{\alpha}\left(\frac{3\pi}{2}\right)^{\frac{2}{3}}
  \left(2l -\frac{3}{4}\right) ^{\frac{2}{3}},
\end{equation}
and for maxima:
\begin{equation}
  \label{eq:Smax}
  s_{max}=\frac{1}{\alpha}\left(\frac{3\pi}{2}\right)^{\frac{2}{3}}
  \left(2l +\frac{1}{4}\right) ^{\frac{2}{3}},
\end{equation}
where $l$ is a non-negative integer that provides us with information
about the number of cycles that have occurred for a particular value
of $l$.

In this way, the distance between two consecutive extremes is thus
given by:
\begin{equation}
  \label{eq:DeltaS}
  \Delta s = \frac{1}{\alpha}\left(\frac{3\pi}{2}\right)^{\frac{2}{3}}\left[
    \left(2l -\frac{3}{4}\right) ^{\frac{2}{3}} -  \left(2l
      +\frac{1}{4}\right) ^{\frac{2}{3}}\right].
\end{equation}
We know that the distance between two consecutive peaks in the beam 
intensity cannot be smaller than a transverse wavelength $\lambda_c$, 
i.e. $\Delta s \geq \lambda_c/2$. This argument provides a physical 
criterion for the cut-off point of a finite Airy beam after which it 
must dampen.

To define the critical transverse wavelength $\lambda_c$ we must also
take into consideration that Airy beams propagate within the paraxial
regime as governed by Equation (\ref{eq:ParaxialWaveEq}). For this purpose we require to provide the maximum angle allowed for a ray in the wavefront to be considered paraxial. Curiously enough, to date in the literature there is not an established quantitative criterion for such angle and, as noticed by Beck several decades ago: ``The term paraxial rays is a relative one and to some extent a matter of arbitrary choice.'' It will depend on the tolerance error that is planned to be accepted \cite{Beck1944}.

We will introduce the quantitative criterion of paraxiality put forward by Agrawal, Siegman and others \cite{Agrawal1979,Siegman1986lasers,Rodriguez-Morales2004a} where paraxial optical beams can be focused or diverge at angles up to a maximum critical value of $\theta_c=\pi/6$. With this in mind, we consider a ray perpendicular to the wavefront of ${\rm Ai}H^{(2)}$ at the cut-off position. When the wavevector $\overrightarrow k$ to this ray reaches the maximum paraxial angle $\theta_c=\pi/6$, recalling that $|\overrightarrow k|=2\pi/\lambda$, a simple calculation gives $\lambda_c = 2\lambda$. Thus, substituting $\Delta s = 2\lambda$ in Equation (\ref{eq:DeltaS}) and solving for $l_c$ we can get the position $s_{max}(l_c)$ where the oscillations of
the Airy profile must dampen.

\subsection{Maximum propagation distance of finite Airy beams}
\label{subsec:propagation}
\begin{figure}[htbp]
  \centering
  \includegraphics[width=8.5cm]{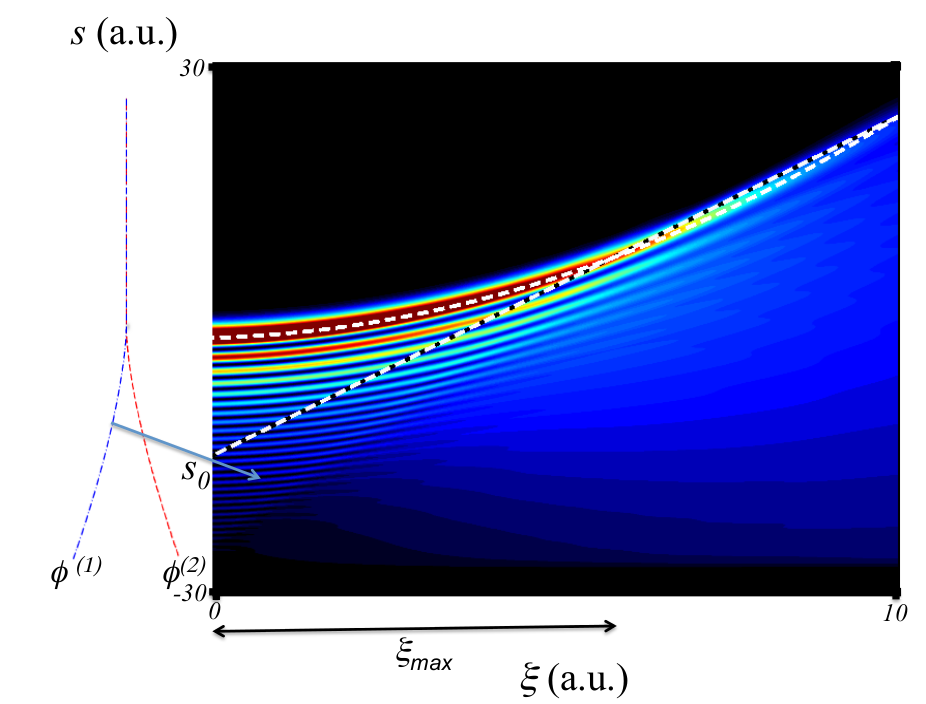}
  \caption{(Color online) The method to determine the propagation
    distance $\xi_{max}$ of an Airy beam is given by the intersection
    of the parabolic trajectory followed by the main maximum and a ray
    coming from the edge of the profile (in arbitrary dimensionless
    units). We also show a diagram of the phases of the Hankel
    components of the Airy beam.}
  \label{fig:Propagation_Distance}
\end{figure}

We now propose a method for determining the propagation distance of
the finite Airy beam based on the observations made above. Consider
the parabolic trajectory followed by the main maximum of the Airy beam
and take a ray coming from the wavefront at the opposite extreme of
the window and find the point of intersection. Figure
\ref{fig:Propagation_Distance} shows this situation; in order to guide
the eye we have marked the region of existence of the Airy beam given
by the ray as well as the parabolic trajectory followed by the beam
(white dotted lines).
 
We know that for an Airy beam propagating in a homogeneous medium, the
normalised solution is given by Equation (\ref{eq:SolAiry}) and thus
it is clear that the trajectory of the main maximum satisfies the
following equation:
\begin{equation}
  \label{eq:trajectory}
  s-\left(\frac{\xi}{2} \right)^2=a_1
\end{equation}
where $a_1=-1.0187297\ldots$ is the first zero of the derivative of
the Airy function, that is to say, the position of the first intensity
maximum at the onset of propagation. Furthermore, we know that light
rays are normal to the wavefront and therefore in this case, the
equation for the ray we are interested in can be expressed as
\begin{equation}
  \label{eq:ray}
  \xi=s_0^{-1/2}(s-s_0),
\end{equation}
where $s_0$ is the point where the beam is truncated, in other words,
where the edge of the window is located.  Using Equation
(\ref{eq:trajectory}) and Equation (\ref{eq:ray}) for the ray we have
that
\begin{eqnarray}
  \label{eq:solution_ray_xi}
  \xi_{\pm} &=& 2\left(\sqrt{|s_0|} \pm \sqrt{2|s_0|-|a_1|} \right),\\
  \label{eq:solution_ray_s}
  s_\pm &=& 3s_0 \pm 2\sqrt{|s_0|}\sqrt{2|s_0| -|a_1|}.
\end{eqnarray}
The solution we seek is the one with the negative sign because it is
the first intersection between the parabola and the ray, i.e. the
point where the propagation ends. In this way, the propagation
distance is given by:
\begin{equation}
  \label{eq:propagation_distance}
  \xi_{max}=2\left(\sqrt{|s_0|} - \sqrt{2|s_0|-|a_1|} \right).
\end{equation}
This expression defines the maximum propagation distance for a finite
Airy beam. This is, to our knowledge, the first time that the
propagation distance for Airy beams is formally defined.

\section{Higher-Order Airy Modes in GRIN media}
\label{sec:HigherOrder}
Up to now we have concentrated on the treatment of the Airy beam
propagating in homogeneous media as a result of solving the normalised
paraxial wave equation given by expression (\ref{eq:ParaxialWaveEq})
and ended up with a solution given by the Airy function. However, the
Airy functions also appear in the description of the propagating modes
in inhomogeneous media, in particular in one dimensional linear
gradient index media \cite{Marcuse1973,chen,Chavez11}.

It is known that in sol-gel materials the gradient-index profile can
be controlled by diffusion and precipitation of dopants, including
quadratic, quartic and higher gradient profiles
\cite{Moore1,Moore2,Moore1995}.  This materials are used to construct
Wood lenses that are thin radial gradient index lenses with plane
parallel surfaces whose index of refraction is of the form
$N=N_{00}+N_{10}r^2+N_{20}r^4+...$ With this antecedent, it can be
possible to create a material whose index of refraction be the one
dimensional equivalent of a Wood lens.

In this Section we investigate the propagating modes for power law
gradient index media and show that they are what we have come to call
higher-order Airy beams.

Let us start with the Helmholtz equation in (1+1)D
\begin{equation}
  \label{eq:HelmholtzHG}
  \nabla^2 E +k^2E=0,
\end{equation}
and let the medium have a dispersion relation such that the wave
number can be expressed as $k^2(x)=k_0^2 +\beta \left[- k_0 -
  x^n+\frac{\beta}{4}\right]$, where $\beta$ is a parameter that
depends on the medium. We can now propose an Ansatz such that the
electric field is given by
$E(x,z)=U(x,z)\exp\left[i\left(k_0-\frac{\beta}{2}\right)z
\right]$. In the paraxial approximation we have that
\begin{equation}
  \label{eq:1_1Simp}
  2 i \left(k_0
    -\frac{\beta}{2}\right) \frac{\partial U}{\partial z}
  +\frac{\partial^2 U}{\partial x^2}-\beta 
  x^n U=0.
\end{equation}
If we require that $\beta=2 k_0$ the wave number is given by
$k^2(x)=-2k_0x^n$ and Equation (\ref{eq:1_1Simp}) is simplified into
\begin{equation}
  \label{eq:ModifiedAiry}
  \frac{\partial^2 U}{\partial x^2}-\beta x^n U=0,
\end{equation}
which corresponds to a generalised form of the Airy differential
equation. Making the transformation \cite{Jahnke43}
\begin{equation}
  \label{eq:U_Bessel}
  U(x) =\sqrt{x} Z_{\frac{1}{n+2}}\left(\frac{2\sqrt{\beta}}{n+2}
    x^{\frac{n+2}{2}} \right),
\end{equation}
substituting into Equation (\ref{eq:ModifiedAiry}) and after some
algebra, yields to
\begin{equation}
  \label{eq:BesselDE_Fractional}
  Z''_{\frac{1}{n+2}}(\zeta) + \frac{1}{\zeta} Z'_{\frac{1}{n+2}}(\zeta) +
  \left(1- \frac{1}{(n+2)^2\zeta^2} \right) Z_{\frac{1}{n+2}}(\zeta) =0  
\end{equation}
where $\zeta= \frac{2\sqrt{\beta}}{n+2} x^{\frac{n+2}{2}}$. Equation
(\ref{eq:BesselDE_Fractional}) is the Bessel differential equation
whose solutions are the family of Bessel functions of order
$\frac{1}{n+2}$.

Equation (\ref{eq:ModifiedAiry}) has been studied in detail by Swanson
and Headley \cite{Swanson1967} who defined its solutions as $A_n(x)$
and $B_n(x)$. It is clear that when $n=1$ the functions $A_n(x)$ and
$B_n(x)$ become the standard Airy functions $\Ai{x}$ and $\Bi{x}$,
respectively. An important remark is that they found that the general
behaviour of the solutions is different depending on the parity of
$n$. This is something to be expected since the branches of the power
law parabolae can have different signs depending on which side of the
origin they extend to. 

We can now refer to Equation (\ref{eq:ModifiedAiry}) as the Airy
differential equation of order $n$. In other words, its solutions can
be seen as higher order Airy functions, and similarly to the Airy
functions, they can also be cast in terms of Bessel functions of
fractional order. It should therefore be clear that given the
adequate conditions in a power law gradient index medium it is then possible to obtain higher-order Airy beams.

We want to remark that the study presented in this section can also be applied to investigate the homologous problem in Quantum Mechanics for the
Schr\"odinger equation of a particle confined within an infinite wall and a power law potential $V(x) = \beta x^n$, with $\beta$ constant
\cite{landau1981quantum,Bernardini,muller2006, santos2005,
  sakurai2011}. For the particular case of the linear potential \cite{miller2008}, the standing Airy wave packet arises from considering a particle subject to a constant force, e.g. gravitational force, that when it takes the value of zero the particle wave packet is still described by the Airy function, but accelerating away from the infinite wall as a
consequence of removing the stabilising force \cite{Chavez11}.

\section{Conclusions}
In this paper we introduced the physical principles that govern the existence and propagation of Airy beams. We have shown that the non-diffracting characteristics of Airy beams can be explained under the formalism of travelling Hankel waves originally introduced to describe Bessel beams. This is possible due to the fact that Bessel and Airy functions are intimately
related to each other, with the latter being the Bessel functions of fractional order equal to $\frac{1}{3}$. We introduced the
two Hankel components of the Airy beam, namely $\AiH{1}{\cdot}$ and
$\AiH{2}{\cdot}$, and showed that the later bears the parabolic caustic
property of the beam. It was shown that the superposition of these Hankel components fully explain in simple and straightforward terms propagation characteristics of the Airy beam, such as self-healing and double-focusing. Also, this approach allowed to establish for the first time the needed expression to compute the maximum propagation distance of finte energy Airy beams. We addressed physical constraints of why an ``ideal'' Airy beam of infinite extent cannot exist and provided a quantitative method to obtain the maximum extent of an Airy beam of a given wavelength. And finally, by studying the solution of the paraxial wave equation in power law GRIN media we demonstrated the possibility of creating higher-order Airy beams.

\begin{acknowledgements}
  The authors would like to acknowledge to Prof. Roc{\'\i}o Jauregui
  Renaud for fruitful and motivating discussions and by bringing to
  our attention Ref. \cite{landau1981quantum}. We would also like to
  thank Prof.  Demetrios Christodoulides his valuable comments during
  the development of the present work, and to Prof. Duncan Moore for
  enlightening discussions on GRIN materials. Finally, we also
  acknowledge support of INAOE, Mexico and the STRI, UK during the
  development of this project.
\end{acknowledgements}


\section*{References}

\end{document}